\input harvmac
\input epsf

\Title{
\rightline{hep-th/0211163}}
{\vbox{\centerline{On D-branes in the Nappi-Witten and GMM}
\centerline{gauged WZW models}}}

\medskip
\centerline{Gor Sarkissian\foot{e-mail: gor@ictp.trieste.it}}
\bigskip
\smallskip
\centerline{The Abdus Salam International Centre for Theoretical Physics}
\centerline{Strada Costiera 11, Trieste 34014, Italy}

\smallskip

\bigskip\bigskip\bigskip
\noindent

We construct D-branes in
the Nappi-Witten (NW)  and Guadagnini-Martellini-Mintchev (GMM) 
gauged WZW models. 
For the $SL(2,R)\times SU(2)/U(1)\times U(1)$ NW 
and $SU(2)\times SU(2)/U(1)$ GMM models we present the explicit 
equations describing
the D-brane hypersurfaces in their  target spaces.
In the latter case 
we show that the D-branes are classified according to 
the Cardy theorem.
We also present the semiclassical mass computation 
and find its agreement with the CFT predictions.

\vfill

\Date{11/02}

\lref\malmosei{J. Maldacena, G. Moore and N. Seiberg, 
Geometrical interpretation of D-branes in gauged WZW models,
JHEP 0107 (2001) 046,
hep-th/0105038.}%

\lref\sark{G. Sarkissian, Non-maximally symmetric D-branes
on group manifold in the Lagrangian approach, JHEP 0207 (2002) 033,
hep-th/0205097.}

\lref\sargor{ G. Sarkissian, On DBI action of the non-maximally 
symmetric D-branes on SU(2), hep-th/0211038.}

\lref\cardy{J. L. Cardy, Boundary conditions, fusion rules and the Verlinde
formula, Nucl. Phys. B324 (1989) 581.}%

\lref\walzh{M. A. Walton and J.-G.  Zhou,
D-branes in asymmetrically gauged WZW models and axial-vector 
duality, hep-th/0205161.}

\lref\napwit{C. R. Nappi, E. Witten
A Closed, Expanding Universe in String Theory,
Phys. Lett. B293 (1992) 309-314, hep-th/9206078.}

\lref\quella{ T. Quella, On the hierarchy of symmetry breaking D-branes 
in group manifolds, hep-th/0209157.}

\lref\gumami{ E. Guadagnini, M. Martellini and M. Mintchev,
Scale invariance sigma models on homogeneous spaces, Phys. Lett.
B194 (1987) 69.}

\lref\guadag{E. Guadagnini, Current algebra in sigma models on 
homogeneous spaces, Nucl. Phys. B290 (1987) 417.}

\lref\zayts{ L. A. Pando Zayas, A. A. Tseytlin,
Conformal Sigma Models for a Class of $T^{p,q}$ Spaces,
Class. Quant. Grav. 17 (2000) 5125-5131, hep-th/0007086.}

\lref\klsev{C. Klimcik and P. Severa, 
Open strings and D-branes in WZNW models,
Nucl. Phys. B488 (1997) 653, hep-th/9609112.}%

\lref\alschom{A. Alekseev and V. Schomerus, 
D-branes in the WZW model, Phys. Rev. D60, 061901 (1999),
hep-th/9812193.}%

\lref\gaw{K. Gawedzki, 
Conformal field theory: a case study, in Conformal Field Theory,
Frontier in Physics 102, Eds. Y. Nutku, C. Saclioglu, T. Turgut
( Perseus Publishing, 2000),
hep-th/9904145.}%

 \lref\ngaw{K. Gawedzki, 
Boundary WZW, G/H, G/G and CS theories,
hep-th/0108044.}%

\lref\elsar{S. Elitzur and G. Sarkissian, 
D-branes on a gauged WZW model,
Nucl.Phys. B625 (2002) 166-178,
hep-th/0108142.}

\lref\witt{E. Witten, On holomorphic factorization of WZW and coset models,
Comm. Math. Phys. 144, 189 (1992).}

\lref\elrgk{S. Elitzur, A. Giveon, D. Kutasov, E. Rabinovici,
From Big Bang to Big Crunch and Beyond, JHEP 0206 (2002) 017,
hep-th/0204189.}

\lref\boris{P. Bordalo, S. Ribault and C. Schweigert, 
Flux stabilization in compact groups, JHEP 0110 (2001) 036,
hep-th/0108201.}

\lref\stanold{S. Stanciu, 
A note on D-branes in group manifolds: flux quantization and D0-charge,
JHEP 0010 (2000) 015, hep-th/0006145.}%
\newsec{Introduction}

\lref\elrsar{S. Elitzur, E. Rabinovici and G. Sarkissian,
On Least Action D-Branes, Nucl. Phys. B541 (1999) 246, hep-th/9807161.}

\lref\harmks{J. A. Harvey, S. Kachru, G. Moore, E. Silverstein,
Tension is Dimension, JHEP 0003 (2000) 001, hep-th/9909072.}

In recent years, some progress in the understanding
of branes on the target spaces
of gauged WZW models has been made. Using the Lagrangian
approach to WZW models developed in \klsev\ ,
\alschom\ and \gaw\ , in  \ngaw\ and \elsar\
D-branes in the vectorially gauged WZW models were constructed.

But it was shown in \witt\ that in order to construct 
gauge invariant WZW model it is enough to satisfy the condition
\eqn\ancon{{\rm Tr}(T_{a,L}T_{b,L}-T_{a,R}T_{b,R})=0,}
where $T_{a,L}$ and $T_{a,R}$ are any generators of the left and right
embedding of the gauge group. The vectorial and axial gauging
correspond to the trivial solutions of \ancon\ $T_{a,L}=T_{a,R}$ and
$T_{a,L}=-T_{a,R}$. Consequently it is an intersting problem to
find boundary conditions invariant under gauge transformations
providing the general solution of \ancon\ .

In \sark\ and \walzh\ new boundary conditions were found admitting
axial gauging for an abelian gauge group.
In \sargor\ ,
the DBI action of the D-branes defined by these boundary conditions
was computed for the group $SU(2)$.

Here, using the D-branes found
in those works, and their diagonal embedding
in products of groups, suggested in \quella\ , 
we present boundary conditions
invariant under an asymmetric action of an abelian gauge group used
in the Nappi-Witten \napwit\ and Guadagnini-Martellini-Mintchev 
\gumami\ , \guadag\  models.
Recently these models have attracted a lof of attention.
The Nappi-Witten model was used to construct
cosmological model of the pre big-bang class \elrgk\ . 
The  Guadagnini-Martellini-Mintchev model provides 
an example of the $T^{p,q}$ spaces \zayts\ .
We believe that the study of the D-brane dynamics in these models will
shed additional light on their properties.

This paper is organised in the following way.

In section 2 we review some facts about non-maximally-symmetric D-branes
necessary for further use. 

In section 3 we present D-branes in the Nappi-Witten model,
construct the action with these boundary conditions and check gauge invariance.

In section 4 we study in detail D-branes in the Nappi-Witten model
considered in \elrgk\ and present the explicit equations of the corresponding 
hypersurfaces.

In section 5, in a similar way 
D-branes in the Guadagnini-Martellini-Mintchev model are considered .

In section 6 we consider in detail D-branes in the $SU(2)\times SU(2)/U(1)$
GMM model. We show that D-branes are classified according to the Cardy theorem. 
We also present a semiclassical mass computation and check
its agreement with the CFT prediction.

\newsec{D-branes on group}
In this section we briefly 
review for further use the results of \sark\ and \sargor\
on the non-maximally symmetric D-branes on group manifolds.

It was shown in \klsev\ that in order to have a well-defined
Lagrangian action of the WZW theory on a world-sheet with boundary,
the boundary conditions should satisfy the following two requirements: 
\item 1. The restriction of the WZW three-form to the D-branes
defined by the boundary conditions should belong to a trivial cohomology
class, i.e. there should exist a two-form $\omega^{(2)}$ satisfying the equation
\eqn\triv{\omega^{\rm WZ}(g)_{|{\rm brane}}
=d\omega^{(2)}.}

It was shown that given a two-form satisfying \triv\
the action can be written in the form
\eqn\actwzw{
S=S(g,k)-\int_D \omega^{(2)},}
where  
\eqn\wzact{S(g,k) ={ k\over{4 \pi}}\left[\int_{\Sigma}d^2z L^{\rm kin} 
+ \int_B \omega^{\rm WZ}\right]}
is the usual WZW action, $D$ is an auxiliary
disc joined to $\Sigma$ along the boundary, completing it to
the closed manifold, and  $B$ is a three-manifold satisfying
the condition $\partial B=\Sigma +D$.
It was understood in \boris\ and \stanold\ that the two-form  $\omega^{(2)}$
is equal to the antisymmetric part of the matrix giving the DBI action:
\eqn\actdb{S_{\rm DBI}=\int\sqrt{{\rm det}(G+\omega^{(2)})},}
or $\omega^{(2)}=B+F$.

\item 2. Some global topological restrictions may arise from
the requirement of the independence of the action \actwzw\
from the actual position of the embedding of the auxiliary
disk in the group manifold. We don't discuss here these conditions,
just stating the results. The details can be found, for example, in \gaw\ .
 
It was found in \alschom\ , 
that maximally-symmetric solutions to these conditions
are quantized conjugacy classes : $C=hfh^{-1}$.
The mentioned two-form was found and has the form
\eqn\twofo{\omega^f(h)=h^{-1}dhfh^{-1}dhf^{-1}.}
Global topological restrictions demand that 
for compact groups $f$ is quantized
and equals 
\eqn\quant{f=e^{i{2\pi\over k}\Lambda\cdot H},} 
where $\Lambda$ are the heighest
weights and $H$ are the Cartan algebra generators.
For example, in the case of the $SU(2)$ group important for us
the branes are given by the quantized set
\eqn\suset{f=e^{i\hat{\psi}\sigma_3},}
where 
\eqn\psihat{\hat{\psi}={2\pi j\over k},\; j=0\ldots {k\over 2}.} 
These D-branes correspond to the Cardy boundary states of the corresponding
CFT model \cardy\ .
In \malmosei\ the non-maximally symmetric
set of boundary states was found, and in \sark\ their
description in the Lagrangian formalism was suggested.
It was shown that the corresponding D-branes can be defined as the product
of a $U(1)$ subgroup and a conjugacy class $g_{\rm boundary}=mC=mhfh^{-1}$,
where $m\in U(1)$. The corresponding two-form was found to be
\eqn\omfor{\omega^{(2)}(m,h)=\omega^{f}(l)-{\rm Tr}(m^{-1}dmdCC^{-1}).}
The topological restrictions demand that conjugacy classes belong
to the same quantized set \quant\ .
In section 6 we will need some details about the non-maximally
symmetric D-branes on SU(2).
In \sargor\ , the two-form \omfor\ was computed
for the case of branes on SU(2). In the Euler
angle parametrisation 
\eqn\euang{g=e^{i\chi{\sigma_3\over 2}}
e^{i\tilde{\theta}{\sigma_1\over 2}}
e^{i\varphi{\sigma_3 \over 2}},}
it was shown that
\eqn\eustf{
 \omega^2(m,h)=B+F={\cos\hat{\psi}\tan{\tilde{\theta}\over 2}\over 
\sqrt{\cos^2{\tilde{\theta}\over 2}-\cos^2\hat{\psi}}}d\tilde{\theta}
\wedge (d\chi-d\varphi)-2\sin^2{\tilde{\theta}\over 2}d\chi\wedge d\varphi.}

\newsec{ D-branes in the Nappi-Witten model}
Let us consider the gauged WZW model $G/H$
defined in the following way \napwit\ .
One takes $G=G_1\times G_2$ and chooses
two U(1) subgroups $U(1)_1\in G_1$ and $U(1)_2\in G_2$.
As gauge group $H$ one takes a 
product of the two U(1) groups, parametrized by 
$\rho$ and $\tau$, $H=U(1)_{\rho}\times U(1)_{\tau}$,
with embeddings 
${\rm em}_{\rho,1}\ :\ U(1)_{\rho}\rightarrow U(1)_1$,
${\rm em}_{\rho,2}\ :\ U(1)_{\rho}\rightarrow U(1)_2$,
${\rm em}_{\tau,1}\ :\ U(1)_{\tau}\rightarrow U(1)_1$,
${\rm em}_{\tau,2}\ :\ U(1)_{\tau}\rightarrow U(1)_2$.
We assume that $U(1)_1$ is generated by $a_1$, 
$U(1)_1=e^{i\lambda_1 a_1}$ and $U(1)_2$ by $a_2$: $U(1)_2=e^{i\lambda_2 a_2}$
and the generators are normalized in the usual way, 
${\rm Tr}a_1^2={\rm Tr}a_2^2=2$.
The action of $H$ we take in the form
\eqn\actgau{(g_1, g_2)\rightarrow 
(h_1g_1h_2,h'_2g_2h'_1)),}
where 
\eqn\embd{\eqalign{
&h_1={\rm em}_{\rho,1}(h_{\rho})=e^{ip\rho a_1},\cr
&h'_1={\rm em}_{\rho,2}(h_{\rho})=e^{iq\rho a_2},}}

$h_{\rho}\in U(1)_{\rho}$,
and
\eqn\embds{\eqalign{ 
&h_2={\rm em}_{\tau,1}(h_{\tau})=e^{ip\tau a_1},\cr
&h'_2={\rm em}_{\tau,2}(h_{\tau})=e^{iq\tau a_2},}}
where
$h_{\tau}\in U(1)_{\tau}$.

The action of the model in the absence of a boundary is
\eqn\actmod{S=S(g_1,k_1)+S(g_2,k_2)+S(g_1,g_2,A_1,A_2),}
where $S(g_i,k_i)$, $i=1,2$ are the usual WZW actions given by \wzact\
and $S(g_1,g_2,A_1,A_2)$ makes the action gauge invariant.
Its explicit form is not important here for us and can be found in \napwit\ .
For gauge invariance,  the levels $k_1$, $k_2$, and
embedding coefficients $p$, $q$ should satisfy
\eqn\coefga{k_1p^2=k_2q^2.}
Now we consider the model in the presence of a boundary.
We take the $U(1)_{\alpha}$ group parametrized by $\alpha$
and consider embeddings $em_{\alpha, 1}\ :\  U(1)_{\alpha}
\rightarrow U(1)_1$, and $em_{\alpha, 2}\ :\  U(1)_{\alpha}
\rightarrow U(1)_2$.
We define the boundary conditions
\eqn\nwbc{g=(g_1, g_2)|_{\rm boundary}=(m_1C_1,m_2C_2),}
where 
\eqn\embdb{\eqalign{
&m_1=em_{\alpha, 1}(m_{\alpha})=e^{ip(\alpha+\gamma_1) a_1},\cr
&m_2=em_{\alpha, 2}(m_{\alpha})=e^{iq(\alpha+\gamma_2) a_2},}}

and $m_{\alpha}\in U(1)_{\alpha}$, $C_1=l_1f_1l_1^{-1}$
 and $C_2=l_2f_2l_2^{-1}$. The parameters $p$ and $q$ are the same as in 
\embd\ and \embds\ . $\gamma_1$ and $\gamma_2$ are 
possibly quantized \malmosei\ constants.

In other words, we take as the D-branes diagonally embedded
U(1)s multiplied by the conjugacy classes.
These boundary conditions 
were recently suggested in \quella\ . Our description \nwbc\
is slightly different from that in \quella\ ,
and more convenient for present purposes. 

Let us check that the boundary conditions \nwbc\ are invariant
under the gauge transformation \actgau\ :
\eqn\gauinv{\eqalign{& g=(g_1, g_2)\rightarrow (h_1g_1h_2,h'_2g_2h'_1)=
(h_1kl_1f_1l_1^{-1}h_2,h'_2kl_2f_2l_2^{-1}h'_1)=\cr
&((h_1kh_2)(h_2^{-1}l_1)f_1(h_2^{-1}l_1)^{-1},(h'_2kh'_1)(h_1^{'-1}l_2)f_2
(h_1^{'-1}l_2)^{-1}).}} 
We see that the boundary conditions preserve their form under the gauge
transformation, with modified parameters:
\eqn\partran{ \alpha\rightarrow \alpha+\rho+\tau,
\;\; l_1 \rightarrow h_2^{-1}l_1,\;\; 
l_2\rightarrow h_1^{-1}l_2.}
As explained in section 2, in the presence of a boundary the
action should be modified by adding the boundary two-form \sark\ :
\eqn\actmodbb{S=S(g_1,k_1)-{ k_1\over{4 \pi}}\int_D\omega^{(2)}(m_1,l_1)
+S(g_2,k_2)-
{ k_2\over{4 \pi}}\int_D\omega^{(2)}(m_2,l_2)
+S(g_1,g_2,A_1,A_2),}
where
$\omega^{(2)}(m,l)$ is defined by \omfor\ .

We now check that \actmodbb\ is invariant under  \actgau\ accompanied
by \partran\ .
First we compute the change of 
\eqn\bound{S_1=S(g_1,k_1)-{ k_1\over{4 \pi}}\int_D\omega^{(2)}(m_1,l_1)}
under the transformations $g_1\rightarrow h_1g_1 h_2$,
$m_1\rightarrow h_1m_1h_2$ and $l_1\rightarrow h_2^{-1}l_1$,
resulting from the presence of the boundary.
From the Polyakov-Wiegmann identity we get
\eqn\boundch{\eqalign{\Delta^{\rm bound}&S(g_1,k_1)=
-{ k_1\over{4 \pi}}\int_D{\rm Tr}(h_1^{-1}dh_1dm_1m_1^{-1}+
h_1^{-1}dh_1dC_1C_1^{-1}
+h_1^{-1}dh_1C_1dh_2h_2^{-1}C_1^{-1}\cr
&+C_1^{-1}m_1^{-1}dm_1C_1dh_2h_2^{-1}+
C_1^{-1}dC_1dh_2h_2^{-1}).}}
Then we have
\eqn\forch{\Delta\omega^{f}(l)=
{\rm Tr}(dh_2h_2^{-1}C_1^{-1}dC_1+dh_2h_2^{-1}dC_1C_1^{-1}+
dh_2h_2^{-1}C_1dh_2h_2^{-1}C_1^{-1}),}
and
\eqn\changad{\eqalign{&\Delta({\rm Tr}(m_1^{-1}dm_1dC_1C_1^{-1}))
={\rm Tr}(-h_1^{-1}dh_1h_2^{-1}dh_2+h_1^{-1}dh_1dC_1C_1^{-1}\cr
&+h_1^{-1}dh_1C_1dh_2h_2^{-1}C_1^{-1}
-m_1^{-1}dm_1h_2^{-1}dh_2+m_1^{-1}dm_1C_1dh_2h_2^{-1}C_1^{-1}\cr
&+h_2^{-1}dh_2dC_1C_1^{-1}+h_2^{-1}dh_2C_1dh_2h_2^{-1}C_1^{-1}).}}
Collecting \boundch\ ,\forch\ and \changad\ we obtain
\eqn\chtot{\Delta^{\rm bound} S_1={ k_1\over{4 \pi}}\int_D 
{\rm Tr}(h_2^{-1}dh_2m_1^{-1}dm_1-h_1^{-1}dh_1dm_1m_1^{-1}
-h_1^{-1}dh_1h_2^{-1}dh_2).}

Similarly for $S_2$ we obtain
\eqn\chtot{\Delta^{\rm bound} S_2={ k_2\over{4 \pi}}\int_D 
{\rm Tr}(h_1^{'-1}dh'_1m_2^{-1}dm_2-h_2^{'-1}dh'_2dm_2m_2^{-1}
-h_2^{'-1}dh'_2h_1^{'-1}dh'_1).}
Taking into account \embd\ , \embds\ ,\embdb\ and \coefga\ we find
that $\Delta^{\rm bound} S_1+\Delta^{\rm bound} S_2=0$,
proving the gauge invariance of the action \actmodbb\ .

\newsec{$SL(2,R)\times SU(2)/U(1)\times U(1)$ NW model}

Let us consider the $SL(2,R)\times SU(2)/U(1)\times U(1)$ Nappi-Witten model.

Here $G_1=SL(2,R)$, $G_2=SU(2)$, $k_1=-k_2$, $p=-i$, $q=1$,
and the $U(1)_{\rho}\times U(1)_{\tau}$
gauge group acts in the following way:
\eqn\nwgtr{(g_1,g_2)\rightarrow(e^{\rho\sigma_3}g_1e^{\tau\sigma_3},
e^{i\tau\sigma_3}g_2e^{i\rho\sigma_3}).}
The D-branes proposed in section 3 have the form
\eqn\dbrfr{g_{\rm boundary}=(e^{(\alpha+\gamma_1)\sigma_3}C_1,
e^{i(\alpha+\gamma_2)\sigma_3}C_2),}
where $C_1=l_1f_1l_1^{-1}$ and $C_2=l_2f_2l_2^{-1}$ are conjugacy classes,
and $f_2$ belongs to the set \suset.
$\gamma_1$ and $\gamma_2$ are possibly quantized constants.
Now we describe this hypersurface in detail.
For this purpose we introduce Euler angles for $SL(2,R)$ and $SU(2)$,
\eqn\euslp{g_1=e^{\chi_1{\sigma_3\over 2}}e^{i\theta_1{\sigma_2\over 2}}
e^{\phi_1{\sigma_3 \over 2}},}
\eqn\euslpp{g_1=e^{\chi_1{\sigma_3\over 2}}e^{\tau_1{\sigma_1\over 2}}
e^{\phi_1{\sigma_3 \over 2}},}
\eqn\eusu{g_2=e^{i\chi_2{\sigma_3\over 2}}
e^{i\tilde{\theta}_2{\sigma_1\over 2}}
e^{i\phi_2{\sigma_3 \over 2}},}
where the first two formulae describe different patches of $SL(2,R)$
and the last one is the usual Euler parametrisation for $SU(2)$.
It is shown in \sark\ that in the Euler angle parametrisations the
product of a $U(1)$ subgroup and a conjugacy class can be described
by inequalities: $e^{\alpha\sigma_3}C_1$ in the patch given 
by \euslp\ is described by the condition 
\eqn\cosle{\cos{\theta_1\over 2}\leq {{\rm Tr}f_1\over 2},}
and in the patch \euslpp\ by the condition
\eqn\coshle{\cosh{\tau_1\over 2}\leq {{\rm Tr}f_1\over 2},}
and $e^{i\alpha\sigma_3}C_2$ in the parametrisation \eusu\
is given by the condition
\eqn\cosleu{\cos{\tilde{\theta}_1\over 2}\geq {{\rm Tr}f_2\over 2}.}
In order to find the equation of the D-brane hypersurface
we should find $\alpha$ on the $SL(2,R)$ and $SU(2)$ sides
and equate them to each other.
It is easy to find the angle $\alpha$ in each case.
Writing the boundary condition in the form $e^{-\alpha\sigma_3}g_1=C_1$
and taking the trace on both sides we easily obtain in the first patch:
\eqn\alphslp{\cosh(\alpha+\gamma_1-{\chi_1+\phi_1\over 2})={{\rm Tr}f_1\over 
2\cos{\theta_1\over 2}},}
 in the second patch:
\eqn\alphslpp{\cosh(\alpha+\gamma_1-{\chi_1+\phi_1\over 2})={{\rm Tr}f_1\over 
2\cosh{\tau_1\over 2}},}
and for $SU(2)$:
\eqn\alphsupp{\cos(\alpha+\gamma_2-{\chi_2+\phi_2\over 2})={{\rm Tr}f_2\over 
2\cos{\tilde{\theta}_2\over 2}}.}
We see that the conditions \cosle\ , \coshle\ and \cosleu\
are necessary for the existence of solutions to eq. \alphslp\ ,\alphslpp\ 
and \alphsupp\ respectively.
Now using gauge fixing conditions $\chi_1=0$ and $\phi_1=0$
we can explicitly write down the D-brane hypersurface equation.
In the first patch we have
\eqn\eqdbr{\cosh\left(\arccos\left({{\rm Tr}f_2\over 
2\cos{\tilde{\theta}_2\over 2}}\right)+{\chi_2+\phi_2\over 2}+
\gamma_2-\gamma_1\right)=
{{\rm Tr}f_1\over 2\cos{\theta_1\over 2}},}
and in the second patch
\eqn\eqdbrp{\cosh\left(\arccos\left({{\rm Tr}f_2\over 
2\cos{\tilde{\theta}_2\over 2}}\right)+{\chi_2+\phi_2\over 2}+
\gamma_2-\gamma_1\right)=
{{\rm Tr}f_1\over 2\cosh{\tau_1\over 2}}.}

\newsec{D-branes in the Guadagnini-Martellini-Mintchev Model}
We begin by reviewing the model
introduced in \gumami\ and \guadag\ (see also \zayts\ ).
This model is a kind of gauged WZW model based on a group $G_1\times G_2$.
The gauge group $H$ acts in the folowing way:
we choose subgroups $H_1\in G_1$ and $H_2\in G_2$ and take
embeddings $em_{1}\ : \ H\rightarrow H_1$ and $ em_{2}\ : \ H\rightarrow H_2$.
It is assumed that $ H_1$ and $H_2$ are the same subgroups of 
$G_1$ and $G_2$ :$H=H_1=H_2$.
The group $H$ acts by the formula 
\eqn\gauactm{(g_1, g_2)\rightarrow (g_1em_{1}(h^{-1}), em_{2}(h)g_2).}
It was shown in  \gumami\ that the following action is invariant under
\gauactm\ :
\eqn\gmmact{S_{\rm GMM}=S(g_1,k_1)+S(g_2,k_2)+S_{\rm int}(g_1,g_2,k),}
where $S(g_i,k_i)$, $i=1,2$ are the usual WZW actions \wzact\ and 
\eqn\actint{\eqalign{&S_{\rm int}(g_1,g_2,k)=-{k\over {2\pi}}\int d^2 x
({\rm Tr}(R_{\alpha}g_1^{-1}\partial_{\mu}g_1){\rm Tr}(R'_{\alpha}
 \partial^{\mu}g_2g_2^{-1})\cr
&+\epsilon^{\mu\nu}{\rm Tr}(R_{\alpha}g_1^{-1}\partial_{\mu}g_1){\rm Tr}
(R'_{\alpha}\partial_{\nu}g_2g_2^{-1})).}}
Here $R_{\alpha}$ and $ R'_{\alpha}$ are the generators
of the Lie algebra of the subgroup $H$ in $G_1$ and $G_2$ respectively.
It is shown in \gumami\ that for gauge invariance the coefficients
entering in \gmmact\ should satisfy
\eqn\coeff{ k_1=kr',\;\;\; k_2=kr,}
where $r$ and $r'$ are given by the embeddings:
\eqn\embd{
{\rm Tr}(R_{\alpha}R_{\beta})=r\delta_{\alpha\beta},
\;\;\;{\rm Tr}(R'_{\alpha}R'_{\beta})=r'\delta_{\alpha\beta}.}
The conformal field theory defined by this sigma model was discussed
in \guadag\ , where the current algebra and the Virasoro algebra
with a central charge value coinciding with that of the GKO construction
for the coset $(G_1\times G_2)/H$ were found.

Here we consider the case when the gauge group is an abelian
group, parametrized by $\rho$: $H=U(1)_{\rho}$.
As before we assume that $H_1$ is generated by a generator $a_1$, 
$H_1=e^{i\lambda_1 a_1}$ and $H_2$ by $a_2$: $H_2=e^{i\lambda_2 a_2}$,
and that the generators are normalized as usual:
${\rm Tr}a_1^2={\rm Tr}a_2^2=2$.
In this case the gauge group acts as
\eqn\actgr{(g_1,g_2)\rightarrow
(g_1h_1,h_2g_2),}
where 
\eqn\gauembd{\eqalign{
&h_1=em_1(h_{\rho}^{-1})=e^{-ip\rho a_1},\cr
&h_2=em_2(h_{\rho})=e^{iq\rho a_2},}}
$h_{\rho}\in U(1)_{\rho}$
and $p$ and $q$ satisfy the relation
\eqn\multco{k_1p^2=k_2q^2.}
Now we consider the model in the presence of a boundary.
We take the $U(1)_{\alpha}$ group parametrized by $\alpha$
and consider embeddings $em_{\alpha, 1}\ :\  U(1)_{\alpha}
\rightarrow U(1)_1$, and $em_{\alpha, 2}\ :\  U(1)_{\alpha}
\rightarrow U(1)_2$.
We suggest the following boundary conditions:
\eqn\boundcond{(g_1,g_2)|_{\rm boundary}=(m_1C_1,m_2C_2),}
where 
\eqn\embdgmm{\eqalign{
&m_1=em_{\alpha,1}(m_{\alpha})=e^{-ip(\alpha+\gamma_1)a_1},\cr
&m_2=em_{\alpha,2}(m_{\alpha})=e^{iq(\alpha+\gamma_2)a_2}}} 
and $m_{\alpha}\in U(1)_{\alpha}$, $C_1=l_1f_1l_1^{-1}$, $C_2=l_2f_2l_2^{-1}$.
The parameters $p$ and $q$ are the same as in \gauembd\ . 
$\gamma_1$ and $\gamma_2$ are possibly quantized \malmosei\ constants.
These boundary conditions are invariant under \gauactm\ :
\eqn\tranbound{\eqalign{&(m_1l_1f_1l_1^{-1},m_2l_2f_2l_2^{-1})\rightarrow 
(m_1l_1f_1l_1^{-1}h_{1},h_2m_2l_2f_2l_2^{-1})\cr
&=((h_{1}m_1)(h_1^{-1}l_1)f_1(h_1^{-1}l_1)^{-1},(h_2m_2)l_2f_2l_2^{-1}).}}
We see that boundary conditions keep the form with modified parameters
\eqn\parch{\alpha\rightarrow \alpha+\rho,\;\; l_1\rightarrow h_1^{-1}l_1.}
In the presence of a boundary we suggest the following action:
\eqn\gmmactb{S_{\rm GMM}=S(g_1,k_1)-
{ k_1\over{4 \pi}}\int_D\omega^{(2)}(m_{1},l_1)+
S(g_2,k_2)-{ k_2\over{4 \pi}}\int_D\omega^{(2)}(m_2,l_2)+
S_{\rm int}(g_1,g_2,k).}
Now we check that the action is invariant under \gauactm\ accompanied by
\parch\ .
From the formula \chtot\ we easily derive the change of $S_1$
and $S_2$ under a gauge transformation,
\eqn\chfir{\Delta^{\rm boundary}S_1={k_1\over {4\pi}}\int_D
{\rm Tr}(h_1^{-1}dh_1dm_1m_1^{-1}),}
\eqn\chsec{\Delta^{\rm boundary}S_2=-{k_2\over {4\pi}}\int_D
{\rm Tr}(h_2^{-1}dh_2dm_2m_2^{-1}).}
which cancel each other as a consequence of the 
conditions \gauembd\ , \embdgmm\ and \multco\ .

\newsec{$SU(2)\times SU(2)/U(1)$ GMM model}

We begin by describing this model following \zayts\ .

The $SU(2)$ group elements are parametrized as 
\eqn\parsuf{\eqalign{&g_1=\exp(i\phi_1\sigma_3)\exp(i\theta_1\sigma_2)
\exp(i\psi_1\sigma_3),\cr 
 &g_2=\exp(i\phi_2\sigma_3)\exp(i\theta_2\sigma_2)
\exp(i\psi_2\sigma_3).}}
The gauge action of the $U(1)$ subgroup is defined by
\eqn\gauactm{\psi_1\rightarrow\psi_1-p\varepsilon(z,\bar{z}),\,\,\,
\phi_2\rightarrow\phi_2+q\varepsilon(z,\bar{z}).}
In the parametrization \parsuf\ the action \gmmact\ is
\eqn\gmmssu{\eqalign{
&S={1\over 4\pi}\int d^2x[k_1(\partial_{\mu}\theta_1\partial^{\mu}\theta_1
+\partial_{\mu}\phi_1\partial^{\mu}\phi_1+
\partial_{\mu}\psi_1\partial^{\mu}\psi_1+\cos(2\theta_1)\partial_{\mu}\phi_1
\partial_{\nu}\psi_1(\eta^{\mu\nu}+\epsilon^{\mu\nu}))\cr
&+k_2\left(\partial_{\mu}\theta_2\partial^{\mu}\theta_2
+\partial_{\mu}\phi_2\partial^{\mu}\phi_2+
\partial_{\mu}\psi_2\partial^{\mu}\psi_2+\cos(2\theta_2)\partial_{\mu}\phi_2
\partial_{\nu}\psi_2(\eta^{\mu\nu}+\epsilon^{\mu\nu})\right)\cr
&+k_3(\cos(2\theta_1)\partial_{\mu}\phi_1+\partial_{\mu}\psi_1)
(\cos(2\theta_2)\partial_{\nu}\psi_2+\partial_{\mu}\phi_2)
(\eta^{\mu\nu}+\epsilon^{\mu\nu})].}}
For the action to be invariant under \gauactm\ one needs to impose the
following algebraic constraints:
\eqn\algcon{k_1p=k_3q,\;\;\; k_2q=k_3p.}
Multiplying these equation we obtain 
\eqn\multco{k_3=\sqrt{k_1k_2},\;\;\; p/q=\sqrt{k_2/k_1}.}
Fixing the gauge by setting $\phi_2=0$ one gets a background whose metric
is of the (non-Einstein) $T^{1,Q}$ type
\eqn\metrqmm{
ds^2=k[d\theta_1^2+\sin^2\theta_1d\phi^2_1+Q^2(d\theta_2^2+
\sin^2\theta_2d\phi^2_2)+(d\psi+\cos\theta_1d\phi_1+Q\cos\theta_2d\phi_2)^2],}
where we have rescaled all variables by $1/2$, renamed $\psi_2\rightarrow
\phi_2$, $\psi_1\rightarrow \psi$ and introduced
\eqn\newcof{Q=p/q=\sqrt{k_2/k_1},\;\;\; k=k_1.}
The background also includes the antisymmetric tensor field
\eqn\antif{
B_{\phi_1\psi}=k\cos\theta_1,\;\; B_{\phi_1\phi_2}=kQ\cos\theta_1\cos\theta_2
,\;\; B_{\phi_2\psi}=-kQ\cos\theta_2.}
Now we are ready to present D-branes in this background and to compute the 
DBI action. 

The D-branes proposed in section 5 have the form
\eqn\dbex{(g_1,g_2)_{\rm boundary}=(e^{-ip(\alpha+\gamma_1)\sigma_3}C_1,
e^{iq(\alpha+\gamma_2)\sigma_3}C_2),}
where $C_1=h_1f_1h_1^{-1}$ and $C_2=h_2f_2h_2^{-1}$ are conjugacy classes,
$f_1=e^{i\hat{\psi}_1\sigma_3}$ and $f_2=e^{i\hat{\psi}_2\sigma_3}$, and
$\hat{\psi}_1$, $\hat{\psi}_2$ belong to the set \psihat\ .
Let us now find the equation describing this hypersurface.
As before, we should find in the parametrization \parsuf\ the angle $\alpha$
and equate both sides. Writing the boundary conditions as 
\eqn\albcf{{\rm Tr}(e^{ip(\alpha+\gamma_1)\sigma_3}g_1)=2\cos\hat{\psi_1},}
\eqn\albcs{{\rm Tr}(e^{-iq(\alpha+\gamma_2)\sigma_3}g_2)=2\cos\hat{\psi_2},} 
from \albcf\ and \albcs\ we obtain
\eqn\albcff{\cos(p(\alpha+\gamma_1)+\phi_1+\psi_1)={\cos\hat{\psi_1}\over
\cos\theta_1},}
and
\eqn\albcss{\cos(-q(\alpha+\gamma_2)+\phi_2+\psi_2)={\cos\hat{\psi_2}\over
\cos\theta_2}.}
Eliminating $\alpha$ from \albcff\ and \albcss\ we get
\eqn\surgmm{{1\over p}\arccos\left({\cos\hat{\psi_1}\over \cos\theta_1}
\right)-{\phi_1+\psi_1\over p}-\gamma_1=-{1\over q}\arccos\left(
{\cos\hat{\psi_2}\over \cos\theta_2}\right)+{\phi_2+\psi_2\over q}-\gamma_2.}
Using now the gauge fixing condition $\phi_2=0$, and rescaling and renaminig
all the variables as before, we get the D-brane hypersurface on this
$T^{1,Q}$ type space,
\eqn\dbrgmm{
\phi_2=2\arccos\left({\cos\hat{\psi_2}\over \cos{\theta_2\over 2}}\right)
+{2\over Q}\arccos\left({\cos\hat{\psi_1}\over \cos{\theta_1\over 2}}\right)
-{\phi_1+\psi\over Q}+2q(\gamma_2-\gamma_1),}
where $Q$ is defined in \newcof\ .
As before $\theta_1$ and $\theta_2$ satisfy the inequalities
\eqn\thuneq{
 \cos{\theta_1\over 2}\geq \cos\hat{\psi_1},\;\;\;
 \cos{\theta_2\over 2}\geq \cos\hat{\psi_2}.}
The presence of the constant term $q(\gamma_2-\gamma_1)$
reflects the invariance of the action \gmmssu\ under the rotations
$\phi_i\rightarrow\phi_i+\beta_i$,  $\psi_i\rightarrow\psi_i+\delta_i$,
 where $\beta_i$ and $\delta_i$
are constant angles, $i=1,2$. But, as noted in \malmosei\ ,
in the gauged WZW models these symmetries are broken to some discrete
subgroups. In the case in question we have 
\eqn\gamquant{\gamma_1={n_1\over k_1p^2},\;\; \gamma_2={n_2\over k_2q^2},}
where $n_1$ and $n_2$ are integers,
and using \multco\ we have for the last part
\eqn\gamq{2q(\gamma_1-\gamma_2)={2n\over qk_2},}
where $n=n_1-n_2$.
We see that the branes \dbrgmm\ are specified by the three parameters
$\hat{\psi_1}$, $\hat{\psi_2}$ and  $n$, in one-to-one
correspondence with the primaries of the corresponding 
GKO coset model $(SU(2)\times SU(2))/U(1)$.
The last piece that we need for the computation of the mass of the 
D-branes is the field-strength.
It can be derived from the field-strength for the corresponding branes on
groups by imposing the gauge fixing condition.
Using the formula  \eustf\ we have:
\eqn\fielstr{F={k\tan{\theta_1\over 2}\cos\hat{\psi_1}\over
\sqrt{\cos^2{\theta_1\over 2}-\cos^2\hat{\psi_1}}}d\theta_1\wedge
(d\phi_1-d\psi)
-{kQ^2\tan{\theta_2\over 2}\cos\hat{\psi_2}\over
\sqrt{\cos^2{\theta_2\over 2}-\cos^2\hat{\psi_2}}}d\theta_2\wedge d\phi_2
-kd\phi_1\wedge d\psi.}
For further use we note that using \dbrgmm\ we can 
compactly re-write \fielstr\ as
\eqn\fielstrr{
F=-kQ\partial_{\theta_1}\phi_2d\theta_1\wedge
(d\phi_1-d\psi)+kQ^2\partial_{\theta_2}\phi_2d\theta_2\wedge d\phi_2-
kd\phi_1\wedge d\psi.}
We turn to the computation of the DBI action
\eqn\dbact{S=\int\sqrt{{\rm det}\phi_2^{*}(g+B+F)},}
where $\phi_2$ is the embedding \dbrgmm\ of the D-brane into the target space.
We first compute the induced metric on the D-brane.
Inserting \dbrgmm\ in \metrqmm\ we obtain the following elements
of the induced metric $G=\phi_2^{*}g$:
\eqn\indmet{\eqalign{
&G_{\theta_1\theta_1}=k+kQ^2(\partial_{\theta_1}\phi_2)^2\;\;\;
 G_{\theta_1\theta_2}=kQ^2\partial_{\theta_1}\phi_2
\partial_{\theta_2}\phi_2\;\;\; 
G_{\theta_1\phi_1}=kQ(\cos\theta_1\cos\theta_2-1)
\partial_{\theta_1}\phi_2 \cr 
&G_{\theta_1\psi}=
kQ\partial_{\theta_1}\phi_2(\cos\theta_2-1)\;\;\;
G_{\theta_2\theta_2}=
kQ^2(1+(\partial_{\theta_2}\phi_2)^2)\;\;\;
G_{\theta_2\psi}=kQ\partial_{\theta_2}\phi_2(\cos\theta_2-1)\cr
& G_{\theta_2\phi_1}=
kQ(\cos\theta_1\cos\theta_2-1)
\partial_{\theta_2}\phi_2 
\;\;\;
G_{\phi_1\phi_1}=2k(1-\cos\theta_1\cos\theta_2)\cr
&G_{\phi_1\psi}= k+k\cos\theta_1-k\cos\theta_2-k\cos\theta_1
\cos\theta_2\;\;\;
G_{\psi\psi}=2k(1-\cos\theta_2).}}
Inserting \dbrgmm\ in \antif\ and \fielstr\
and adding this to \indmet\ we obtain the following matrix 
$\phi_2^{*}(g+B+F)$:
\eqn\gfbmat{\matrix{
k+kQ^2(\partial_{\theta_1}\phi_2)^2&0&-2kQ\partial_{\theta_1}\phi_2&0\cr
2kQ^2\partial_{\theta_1}\phi_2
\partial_{\theta_2}\phi_2& kQ^2(1+(\partial_{\theta_2}\phi_2)^2)&
-2kQ\partial_{\theta_2}\phi_2&-2kQ\partial_{\theta_2}\phi_2\cr
2kQ\partial_{\theta_1}\phi_2\cos\theta_1\cos\theta_2&
2kQ\partial_{\theta_2}\phi_2\cos\theta_1\cos\theta_2&
2k(1-\cos\theta_1\cos\theta_2)&2k\cos\theta_1(1-\cos\theta_2)\cr
2kQ\partial_{\theta_1}\phi_2(\cos\theta_2-1)&2kQ\partial_{\theta_2}\phi_2
\cos\theta_2& 2k(1-\cos\theta_2)&2k(1-\cos\theta_2).\cr}}

Computing the determinant we obtain
\eqn\detere{{\rm det}\phi_2^{*}(g+B+F)=
16k^4Q^2\left(\sin^2{\theta_1\over 2}+Q^2(\partial_{\theta_1}\phi_2)^2
\cos^2{\theta_1\over 2}\right)\left(\sin^2{\theta_2\over 2}+
(\partial_{\theta_2}\phi_2)^2\cos^2{\theta_2\over 2}\right).}
Finally,  expanding the derivatives we get
\eqn\finact{\sqrt{{\rm det}\phi_2^{*}(g+B+F)}=
 4k^2Q{\sin\theta_1\sin\theta_2\over
\sqrt{2(\cos\theta_1-\cos2\hat{\psi_1})}
\sqrt{2(\cos\theta_2-\cos2\hat{\psi_2})}}.}
Now integrating \finact\ we obtain the energy of the brane:
\eqn\energ{
E_{\rm DBI}=\int_0^{2\hat{\psi_1}}d\theta_1\int_0^{2\hat{\psi}_1}d\theta_2
\int_0^{4\pi}d\phi_1\int_0^{\pi}d\psi\sqrt{{\rm det}\phi_2^{*}(g+B+F)}
=64k^2Q\pi^2\sin\hat{\psi_1}\sin\hat{\psi_2}.}
Let us compare this result with the CFT predictions.
In a rational CFT the Cardy boundary states describing the D-branes are
\eqn\cbs{|a\rangle_C=\sum {S_{ab}\over \sqrt{S_{0b}}}|b\rangle\rangle,}
where $a$ and $b$ label primaries, $|b\rangle\rangle$ are the Ishibashi states,
and $S_{ab}$ is the modular-transformation matrix.
The mass of a Cardy state is given by the coefficient of its
$b=0$ Ishibashi component \elrsar\ , \harmks\ :
\eqn\massc{M_C\sim {S_{a0}\over \sqrt{S_{00}}}.}
In the coset $(SU(2)\times SU(2))/U(1)$ the modular-transformation matrix is the
product of the corresponding matrices of all the constituent groups. 
The S-matrix of SU(2) at level $k$ is
\eqn\modtran{S_{ij}=\sqrt{{2\over k+2}}\sin\left({(2i+1)(2j+1)\pi
\over k+2}\right).}
We want to compare the CFT mass prediction with the DBI result in
the semiclassical limit of large $k$. In this limit
\eqn\modlar{{S_{i0}\over \sqrt{S_{00}}}={(2k+4)^{1/4}\over \pi}
\sin\hat{\psi},}
where $\hat{\psi}$ is defined in \psihat\ .
Collecting everything we obtain that according to the CFT computations
in the semiclassical limit the mass of the D-brane specified
by $\hat{\psi_1}$ and $\hat{\psi_2}$ is
\eqn\massdb{M_C\sim \sin\hat{\psi_1}\sin\hat{\psi_2}}
in agreement with the DBI result \energ\ .

\vskip 30pt

{\bf Acknowledgements:}

\vskip 30pt
 
The author thanks M. Blau and S. Elitzur for useful comments.
 
\listrefs
\end